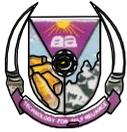

# A FEASIBILITY STUDY OF USING LIGHT FIDELITY WITH MULTIPLE UNMANNED AERIAL VEHICLES (UAVS) FOR INDOOR COLLABORATIVE AND COOPERATIVE NETWORKING

**B.O. Sadiq[1], A.E. Adedokun[2], M.B. Mu'azu[3] and Y. A. Sha'aban[3]**

*Department of Electrical and Computer Engineering, Ahmadu Bello University Zaria, Nigeria.*
*sadiqbashirolaniyi@gmail.com[1], wale@abu.edu.ng[2], mbmuazu@abu.edu.ng[3]*

**ABSTRACT**

This paper evaluates the feasibility of using light fidelity with multiple Unmanned Aerial Vehicles (UAVs) for indoor collaborative and cooperative networking. A number of works have presented challenges for wireless networking with multiple UAVs such as bandwidth efficiency, security and environment amongst others. Due to this fact, there is a need to secure information and data in UAV ad hoc network (FANETs) using a secure transmission medium that guarantees bandwidth efficiency such as the light fidelity (Li-Fi). Li-Fi enables data transmission using visible light through a light emitting diode (LED) and is chosen as against wireless fidelity (Wi- Fi) because it is more ideal for high density wireless data coverage in confined areas, better at addressing radio interference related issues and provides better bandwidth, efficiency, availability and security. Thus, using Light Fidelity as a means of communication necessitates the need to develop a Link Velocity Based Connectivity Algorithm (LVCA) that governs communication between the UAVs and the Li-Fi transmitter and receiver circuits for collaborative and cooperative networking.

## 1.   INTRODUCTION

The use of UAVs is on the increase due to their flexible platforms, low costs and ease of deployment. This makes them a very attractive technology for many civilian indoor and outdoor applications such as animal tracking, environmental and remote sensing, oil pipeline monitoring, traffic surveillance, media coverage and information broadcasting, sports coverage, production processes monitoring amongst others (Bekmezci *et al.*, 2015) in addition to the traditional military and security applications.

When more than one UAV is being flown, and used in a given location for different purposes, these UAVs are called flying UAVs. The communication networks used to share resources between the multiple UAVs with a view to exchanging information are known as flying ad-hoc networks (FANETs) (Bekmezci *et al.*, 2013). A multi-UAV system comprises of a multiple of UAVs whose operations and communications are network-centric i.e. FANET based. However, a fundamental but tasking problem in the collaboration and cooperation of these multiple UAVs is the efficient networking of the UAVs over the unguided medium in rapidly changing environment. A typical multi-UAV system in which the UAVs are communicating with each other as well as with the ground station is as shown in Figure 1.

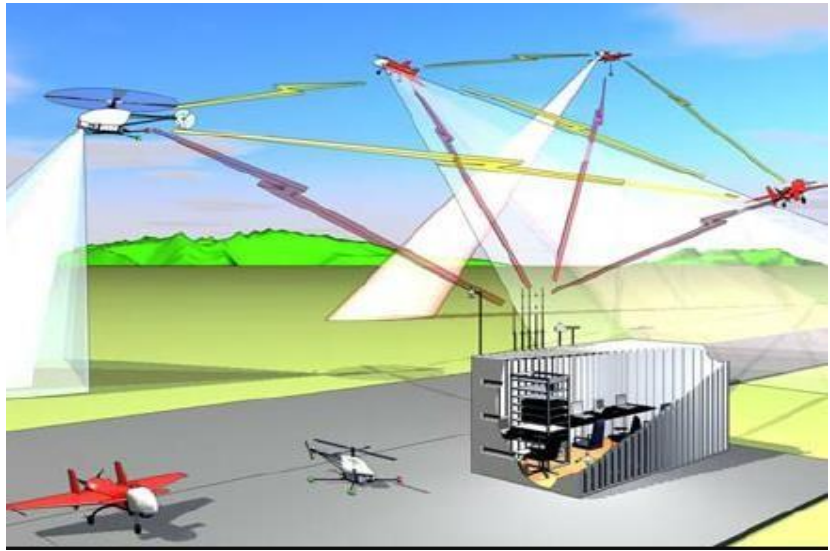

**Figure 1: Typical Multi-UAV System (Göktoğan & Sukkarieh, 2008)**

Multi-UAV systems have a number of advantages over the single UAV systems such as speed of acquiring and disseminating information, Cost, Scalability, Redundancy, Small radar coverage amongst others. Despite these advantages, there are still the encounters of communication channel, power consumption and appropriate communication protocol amongst others (Bekmezci *et al.*, 2013). This implies that designing efficient network architectures that will enable the UAVs to communicate with each other over a wireless channel is a critical issue to consider.

In order to overcome these limitations, a communication solution for multi-UAV systems can be achieved by creating an ad-hoc network between UAVs, which is called FANET. All UAVs in the multi-UAV system constitute the FANET and as such, the UAVs can communicate with each other and the ground base or satellite via the communication channel. The existing wireless communication channel used by earlier researchers was the Wi-Fi (Wireless Fidelity), 3G and satellite. The Wi-Fi however, posed some challenges for indoor application such as security due to the fact that it can penetrate through walls. This made it inefficient in security applications that are mission dependent (Dutta *et al.*, 2013). Li-Fi, an optical version of the Wi-Fi is an emerging technology in wireless communication. Li-Fi technology is based on LEDs for the transfer of data. The transfer of the data can be with the help of different kinds of light, irrespective of the part of the spectrum they belong. That is, the light can belong to the invisible, ultraviolet or the visible part of the spectrum (Pathak *et al.*, 2015). Li-Fi composed of a transmitter and receiver circuits offers a new and unutilized bandwidth of visible light to the currently available radio waves for data transmission. Thus, it offers much larger frequency band (300 THz) as compared to that available in radio frequency (RF) communications (300GHz) (Poonam & Siddiqui., 2014). The Li-Fi is chooses as against the Wi-Fi because:
  i. For indoor surveillance and information sharing, Li-Fi guarantee security
  ii. It provides high speed data that can be used for bandwidth required applications
  iii. The cost of LEDs is cheap

However, Li-Fi technology is not expected to replace the Wi-Fi technology completely. They are expected to be used alongside each other. Therefore, The Use of light fidelity as a medium for networking multi-UAV system that forms an ad hoc network with each other arose as a result of optimally securing data and guaranteeing bandwidth efficiency when UAVs are employed for civilian/ military purposes that are mission specific. Therefore, using the Li-Fi technology with multiple unmanned aerial vehicles necessitates the need to develop a Link Velocity Based Connectivity Algorithm (LVCA) that governs communication between the UAVs and the Li-Fi transmitter and receiver circuits.

The objective of this paper is to explain how multiple UAVs that forms FANET can be networked using Light Fidelity as a medium of communication. This will deal with the communication medium itself and routing issues in FANET with a view to achieving seamless communication.

## 2. STATE OF THE ART

The early researchers have surveyed and identified the concepts, challenging issues and networking models in FANETs using Wireless Fidelity (Wi-Fi) as a medium of communication. Some of the pertinent ones are reviewed in this section.

Sahingoz, (2014) clearly stated the advantages of a multi-UAV system in FANET over that of a single UAV system. In a multi-UAV system, task is shared amongst UAVs which increased the fault tolerance of the system. As such, there should be a communication model for the UAVs to share tasks within each other. However, efficient and secure communication in FANET was still the issue yet to be resolved.

In confined environment, a typical example of physical devices composed of a transmitter and receiver for the purpose of sharing information securely is the Light Fidelity (Li-Fi) as presented in the work of Khairi & Berqia, (2015) and Khare *et al.*, (2016). This concept is an emerging technology in the field of wireless networking and can be employed for the purpose of communication in the wireless ad hoc networks for next generation. However, the concept still pose some challenges such as it usage in only indoor application due to external illumination and the problem of how the receiver will transmit back to the transmitter. Schmid & Corbellini, (2013), Saroha & Mehta, (2014), Karunatilaka *et al.*, (2015), Rashmi *et al.*, (2015) Khare *et al.*, (2016), Vijay & Geetha, (2016) and Jadhav *et al.*, (2017) are amongst the earlier researchers that have demonstrated the applicability of LED-to LED communication networks for information sharing. However, Limited or no work has applied the principle of LED-to-LED communication in FANET. Thus, in order to apply this recent technology to FANET, there is a need to design a protocol that govern the communication between the transmitter and the receiver with respect to the application domain.

Due to the fact that the Multi-UAVs form an adhoc network with each other, a number of researchers have then modified or directly applied the existing VANET and MANET specific routing protocol to multi-UAV system in FANET, such as the OLSR in the work of Singh & Verma, (2014), an extension of the OLSR protocol in the work of Rosati *et al.*, (2014). However, the periodic flooding nature and the frequent transmission of control packets like the Hello message of the protocol normally leads to a large amount of overhead. Thus, making it unsuitable for multi-UAVs in FANET.

Also, a modified MANET protocol for FANET was developed in the work of Qingwen *et al.*, (2015). However, insight into the workings, principles and practice of routing protocol were not presented. An experimental procedure of using ant colony optimization (ACO) algorithm to modify existing MANET specific routing protocol with a view to applying it to FANET was presented in the work of Maistrenko & Alexey, (2016). However, the dynamically changing topology and the rate of high mobility of nodes limited the performance of these algorithms and the overhead cost are quite high.

The Authors in the work of (Gankhuyag *et al.*, 2017) developed a novel predictive routing strategy for flying ad hoc networks. The work proposed the use of a combined omnidirectional and directional transmission scheme in order to overcome the limitations of the existing ad hoc routing protocols that fail in FANET, paying particular attention to increasing the reliability of the routing path using derived expressions. Nonetheless, there is a need to develop a FANET routing protocol with peer to peer communication for cooperative synchronization and collision anticipation.

Therefore, in view of the reviewed pertinent literatures there is a need to develop a Velocity Link Based Connectivity Algorithm (VLCA) that governs communication between the UAVs and the Li-Fi transmitter and receiver circuits for collaborative and cooperative networking.

### 3. RESEARCH PROBLEMS AND DESIGN METHODOLOGY

The existing MANET and VANET routing protocols that made use of the Wi-Fi as a wireless communication medium in confined areas cannot satisfy the FANET routing requirements in multi-UAV systems. This is because they do not address the unique characteristics of the networks such as security, cooperation and collaboration of UAVs in rapidly changing environment. Therefore, in order to achieve efficient networking for communication, cooperation and collaboration among multiple UAVs that guarantees security, provides high speed data (useful for bandwidth demanding applications such as multimedia data), next generation networks should be employed.

An illustration of the design platform, research design and methodology is depicted in Figures 2 and 3 respectively.

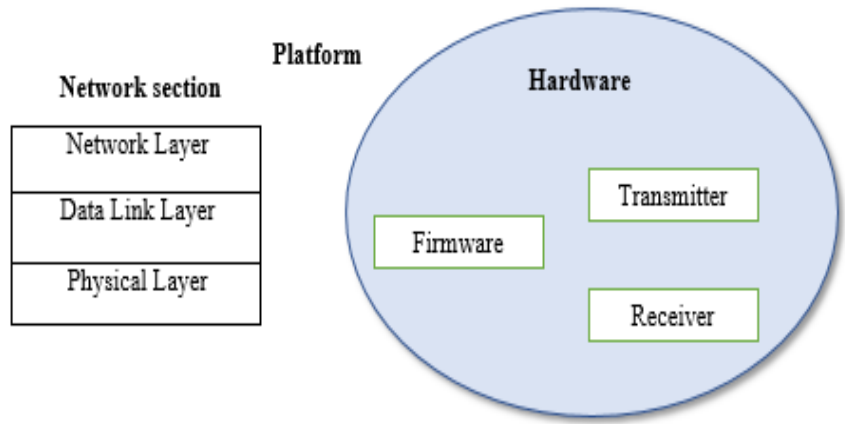

**Figure 2: Design Platform**

### 4. CONCLUSION AND FURTHER WORK

The existing MANET and VANET routing protocols that made use of the Wi-Fi as a wireless communication medium in confined areas cannot satisfy the FANET routing requirements in multi-UAV systems. This is because they do not address the unique characteristics of the networks such as security, cooperation and collaboration of UAVs in rapidly changing environment. Therefore, in order to achieve efficient and seamless networking for communication in FANET, next generation networks should be employed. The current work in progress is the development of a Li-Fi transmitter and Reciever circuit diagrams for establishing the communication link, the Link Velocity Based Connectivity Algorithm (LVCA) and the FANET specific routing protocol.

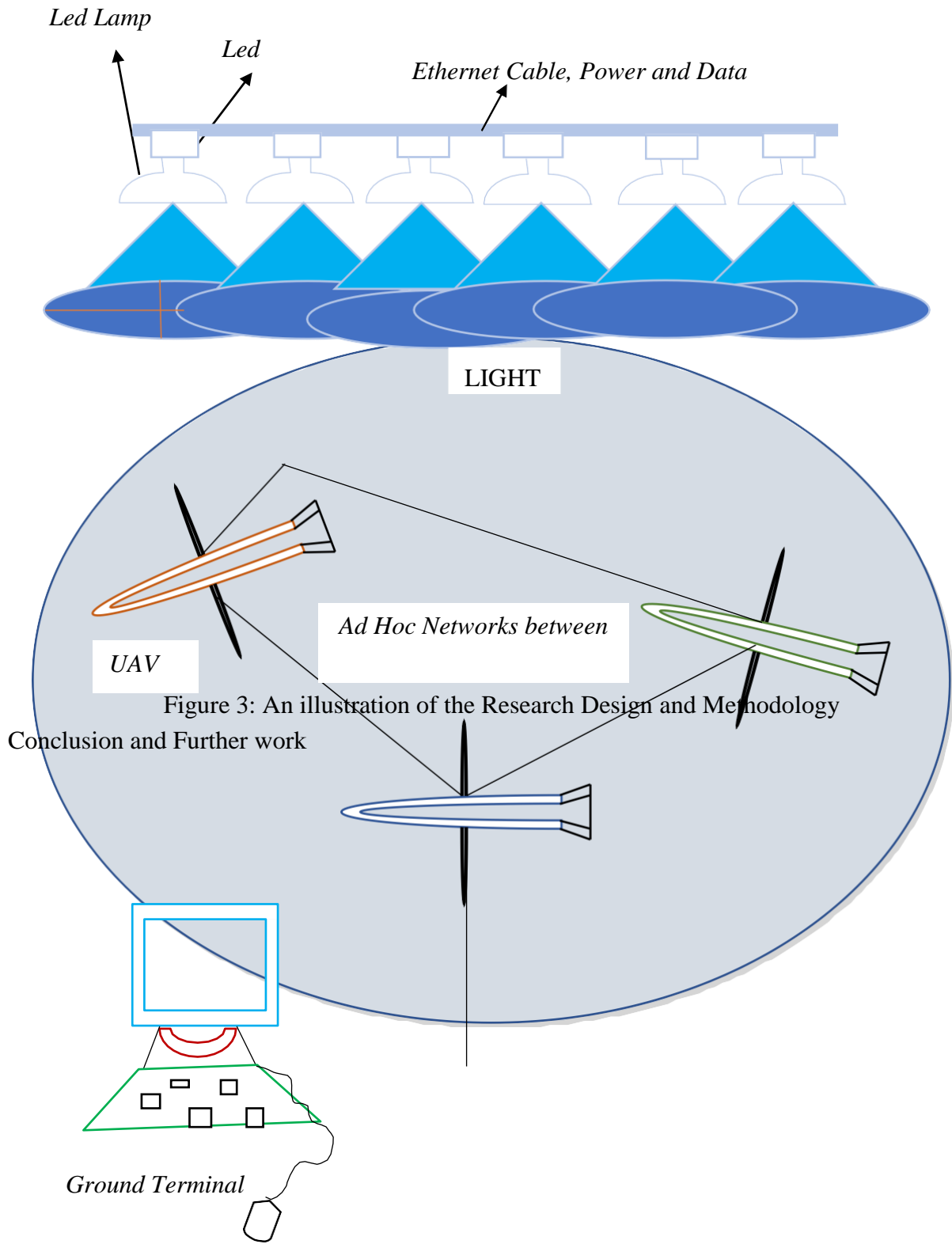

Figure 3: An illustration of the Research Design and Methodology

Conclusion and Further work

**Figure 3: An illustration of the Research Design and Methodology**